\documentclass[preprint,samsmath,amssymb,floatfix,superscriptaddress,nofootinbib,aps]{revtex4}
\usepackage{graphicx}
\begin{document}

\title{Maximum Valency Lattice Gas Models}

  \author{Srikanth Sastry} \affiliation{{Jawaharlal Nehru Centre for Advanced Scientific Research,Jakkur Campus, Bangalore 560064, INDIA  } }
 \author{Emilia La Nave} \affiliation{ {Dipartimento di Fisica and
  INFM-CRS-SOFT, Universit\`a di Roma {\em La Sapienza}, P.le A. Moro
  2, 00185 Roma, Italy} }
\author{Francesco Sciortino} \affiliation{ {Dipartimento di Fisica and
  INFM-CRS-SOFT, Universit\`a di Roma {\em La Sapienza}, P.le A. Moro
  2, 00185 Roma, Italy} } 
         
\begin{abstract}
We study lattice gas models with the imposition of a constraint on the
maximum number of bonds (nearest neighbor interactions) that particles
can participate in. The critical parameters, as well as the
coexistence region are studied using the mean field approximation and
the Bethe-Peierls approximation. We find that the reduction of the
number of interactions suppresses the temperature-density region where
the liquid and gas phases coexist. We confirm these results from
simulations using the histogram reweighting method employing grand
Canonical Monte Carlo simulations. 
\end{abstract}
\maketitle

\section{Introduction}

The study of the gel state of matter in colloidal systems is receiving
significant attention in the last
years\cite{Ver99a,Tra04a,Cip04a,Sci05a}. At the heart of this interest
lies the hope that these studies will help understanding differences
and similarities in the processes of formation of arrested states of
matter at low packing fraction $\phi$, and in particular the
inter-relations between gels and glasses. A similarly ambitious
additional goal is to provide a new route for understanding the
intrinsic properties of the process of formation of physical gels in
systems more complicated than colloids, as the case of gels formed by
reversible cross-linking of polymeric chains\cite{Rub99a} and gelation
in protein solutions\cite{benedek}.  The full comprehension of both
these aggregation processes are of extreme importance in the food
industry, in the protein crystallization process and in the design of
novel biomaterials\cite{Sta05a}.

One of the key questions regards the interplay between the process of
formation of a long-living network and the process of phase separation
in colloid-rich and colloid-poor
regions\cite{Lod99a,bergenholtz,clust,delgado,Zac04b}. Indeed, the
increase of the bond interaction strength (relative to the thermal
energy) controls both the increase of the inter-particle bond lifetime
(and as a consequence the lifetime of the spanning network) and the
increase of the driving force of formation of locally dense packed
states, which progressively favor nucleation of the ``liquid''
(colloid-rich) phase\cite{Sci05a}. As a result, the establishment of
conditions such that a connected percolating structure able to sustain
stress survives for times longer than the observation time is often
preempted by phase separation. For spherical attractive interaction
potentials, it has been found that phase separation is always dominant
at low densities and arrested states at low packing fractions $\phi$
can be reached only in non-homogeneous state, whose morphology is
determined by the phase separation
process\cite{sastry2000,speedy,otp,shell,ashwin2006,Fof05a,manley}.

A possible mechanism to generate low packing fraction arrested states
in the absence of phase separation has been recently proposed and
supported by off-lattice simulations\cite{Zac05b,bianchi2006}.  It has been shown
that the region in the $T-\phi$ plane in which a two-phase coexistence
is thermodynamically preferred as compared to the homogeneous fluid
state can be progressively reduced (both in $T$ and $\phi$) by
decreasing the maximum number of possible pair-wise bonded
interactions.  In an equivalent language, the reduction --- at fixed
interaction potential range --- of the colloid surface allowing for
inter-particle bonding progressively destabilize phase
separation\cite{Sea99a,Ker03a}. According to these ideas, colloidal
particles with a limited number of attractive spots are the best
candidates for gel formation. Interestingly enough, these ideas carry
also to the case of protein solutions, where the character of the
amino acids on the surface of the protein controls the strength and
the directionality of the inter-protein interaction\cite{benedek}.

In this article, we present a lattice gas model, which we solve in the
Bragg-Williams mean field approximation and in the Bethe-Peierls
quasi-chemical approximation and which allow us to visualize in a
clear way the relation between the limitation of the maximum number of
interaction and the amplitude of the phase separated region in the
$T$-$\phi$ phase diagram. We complement these calculations with Monte
Carlo simulations of the liquid-gas phase coexistence, to assess the
reliability of the mean field solutions.  The reported results confirm
that the reduction of the maximum number of interactions is indeed an
efficient mechanism for generating  thermodynamically
stable states at extremely low temperatures.
 
\section{The Model} 

The system we consider is a nearest neighbor lattice gas, with
occupancy variables $n_i$ at each node of a lattice with $\gamma$
nearest neighbors.

The Hamiltonian for the system is written, in the usual form, as 

\begin{equation} 
H = -\epsilon \sum^{*}_{<ij>} n_i  n_j 
\end{equation} 

but with the $^{*}$ indicating that the sum is only over such bond configurations 
that have at each vertex a maximum of $\gamma_m$ bonds.

For the purposes of mean field theory, we can write this as 

\begin{equation} 
H = -{\epsilon \over 2} \sum_i n_i f(\sum_j^{\gamma} n_j) 
\end{equation} 

where the first sum $i$ runs over all lattice sites, and the second
sum $j$ runs over all $\gamma$ nearest neighbors of site
$i$, and with
\begin{eqnarray} 
f(x)  = & x,                   & x \le \gamma_m \\
       = & \gamma_m, & x >  \gamma_m
\end{eqnarray} 

where $\gamma_m$ denotes the maximum number of interactions, or
valency, allowed for a particle at any site. When $\gamma_m = \gamma$,
one recovers the simple lattice gas. In the following, we consider the
behavior of the system when the parameter $\gamma_m$ is varied between
the limits $\gamma_m = 0$ (when one has a non-interacting {\it
paramagnetic} lattice gas) and $\gamma_m = \gamma$ (when one has the
simple nearest neighbor lattice gas).

\section{Mean Field Approximation} 

In order to perform the mean field calculation, we approximate the 
function $f$ as 

\begin{equation} 
f(x) = \gamma_m \tanh(x/\gamma_m) 
\end{equation} 

which has the desired linear behavior for small $x$ and a constant
value of $\gamma_m$ at large $x$. Further, the mean field
approximation amounts to writing $\sum_j^{nn} n_j = \gamma <n>$ where
$<n>$ is the average occupancy, given by $<n> = \sum_i n_i/N$ where $N$ is
the total number of sites. With this approximation, the thermodynamic 
potential $\Omega$ in the grand canonical ensemble may be written as 

\begin{equation} 
\Omega/N  = -{\epsilon \over 2} \gamma_m <n> \tanh({\gamma <n> \over \gamma_m}) - \mu <n> + k_B T \left[ <n> \log <n> + (1 - <n>) \log (1 - <n>)\right] 
\end{equation} 

where $\mu$ is the chemical potential, $T$ is the temperature, $k_B$
is the Boltzmann's constant and $<n>$ is the occupancy, given $\mu$ and
$T$, that minimizes $\Omega$. Imposing the minimization condition, 

$$ {\partial \Omega \over \partial {\bf <n>}} = 0 $$ 

allows one to eliminate $\mu$ and write the thermodynamic potential
$\Omega$ in terms of $T$ and $<n>$. Further, since $\Omega = - P V = - P
N $, (where $P$ is the pressure, and the system volume for the lattice
gas $V = N$ the number of sites) one obtains in this way the equation of
state of the system:

\begin{equation} 
P = - {\epsilon \over 2} \gamma <n>^2 sech^2 ({\gamma <n> \over \gamma_m}) - k_B T \log (1 - <n>)
\end{equation}  

The condition for the critical point is given by 

$$ {\partial P \over \partial <n>} = 0 $$ 

and 

$$ {\partial^2 P \over \partial <n>^2} = 0. $$ 

From these, one obtains the critical density and the critical
temperature, which depend on the parameter $\gamma_m/\gamma$ and are
plotted in Fig. 1. From the figure, it is clear that both the critical
density and the critical temperature go to zero as the variable
$\gamma_m/\gamma$ goes to zero.

\begin{figure}
\includegraphics[width=150mm]{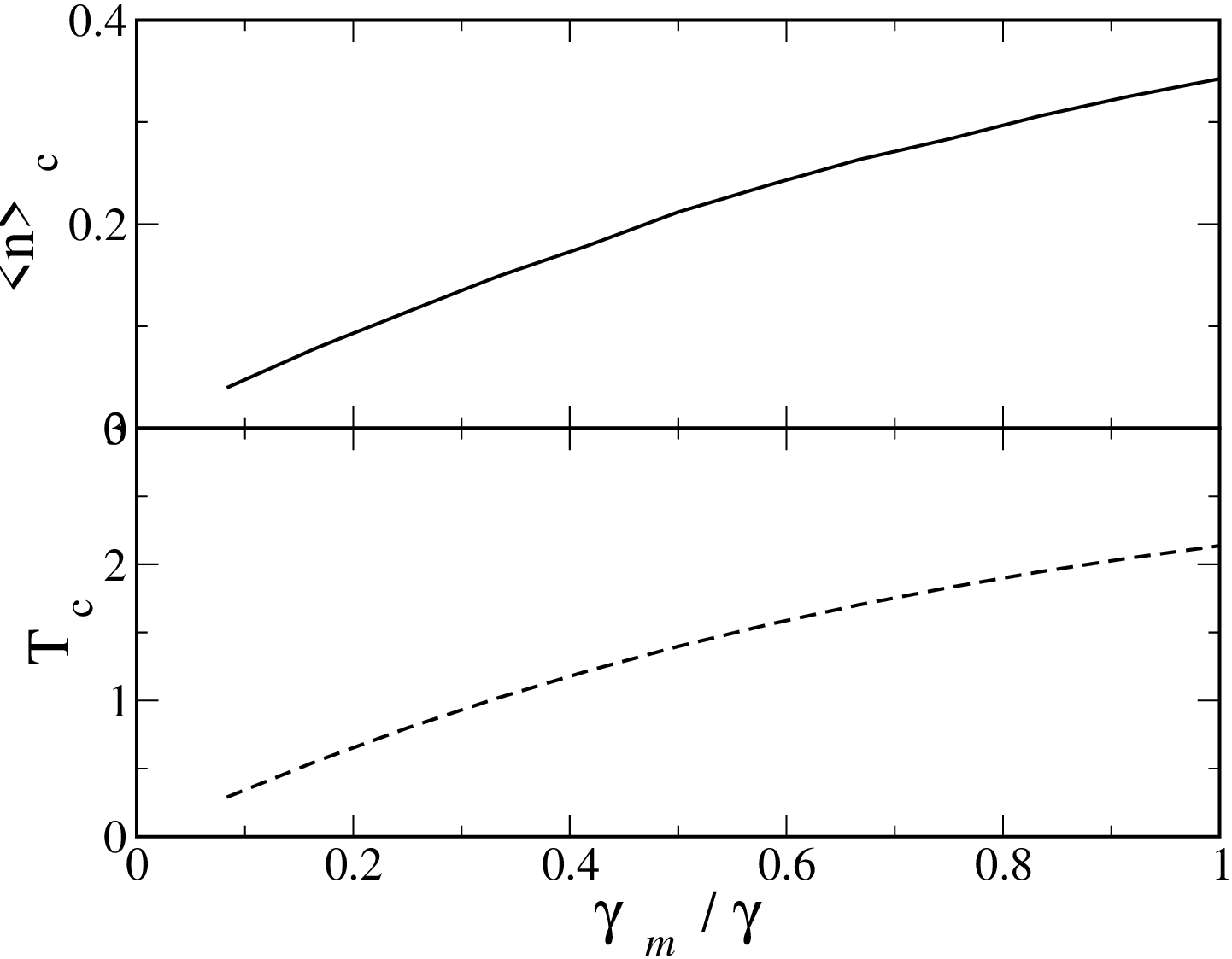}%
\caption{Mean field critical density  (top panel) and critical temperature (lower panel) as a function of parameter  $\gamma_m/\gamma$. Calculated for $\gamma = 12$ and $\epsilon = 1$.}
\label{F_fig1}
\end{figure}

We next calculate the spinodals that demarcate the region of
instability in the ($<n>, T$) and in the ($<n>,P$) phase diagram which are  shown in
Fig. 2.  Consistently with expectations based on the behavior of the
critical density and temperature, the region of instability also shrinks 
with the reduction in the valency.

\begin{figure}
\includegraphics[width=150mm]{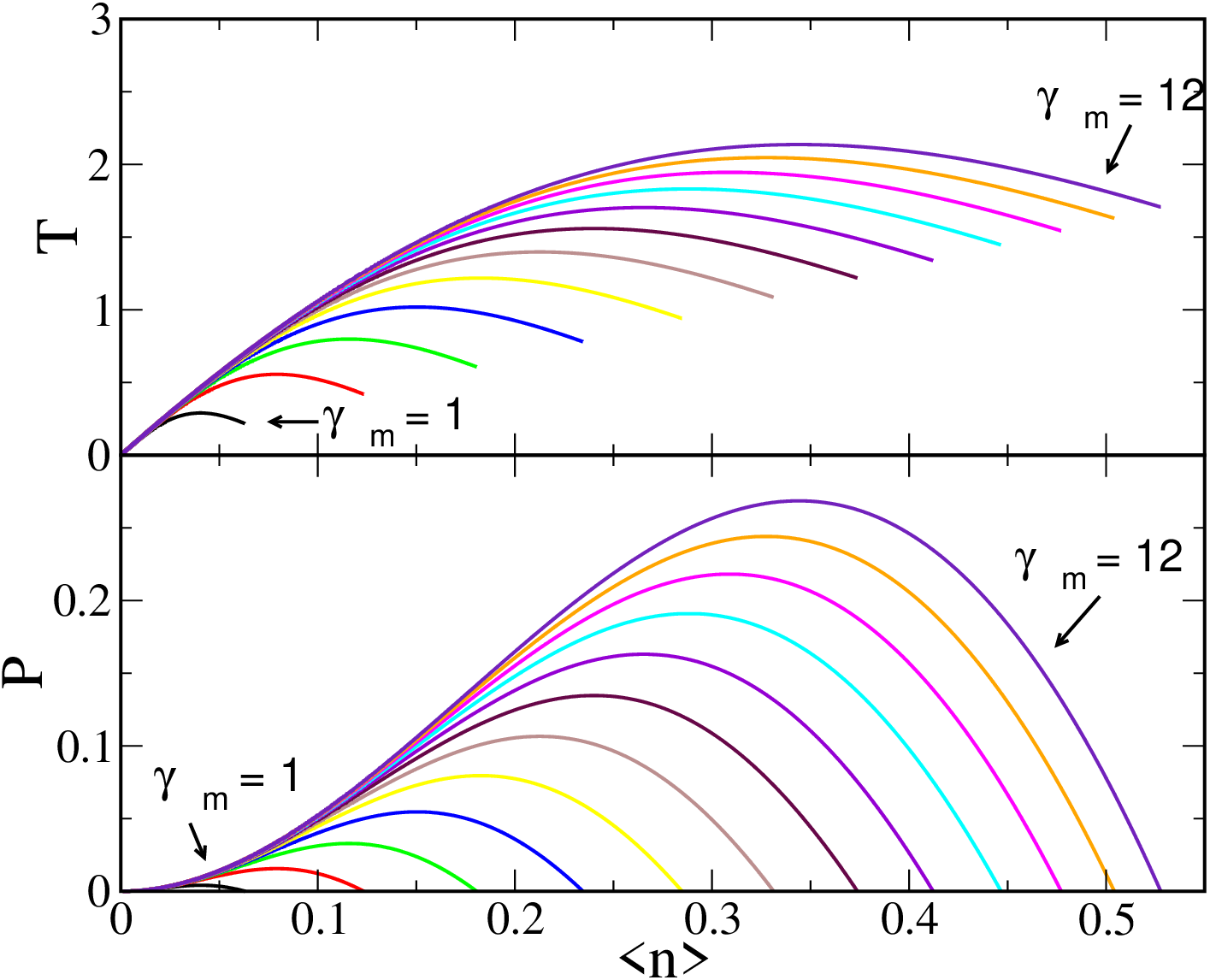}%
\caption{The mean-field spinodals in the ($<n>$, T) phase diagram (top panel) and in the ($<n>$,P) phase diagram
(lower panel)  for different values of the  parameter 
$\gamma_m$. Calculated for $\gamma = 12$ and $\epsilon = 1$.}
\label{F_fig2}
\end{figure}

\section{Bethe-Peierls Approximation}

As an improvement over the mean field approximation, we next consider
the Bethe-Peierls approximation, which in the case of the simple
lattice gas is equivalent to the Bethe lattice. As in the standard
treatment (see, {\it e. g.} \cite{Huang}), we consider a ``center''
site, and its surrounding neighbors. We write the probability that the 
center site is occupied, with $n$ nearest neighbors also being occupied, 
as 

\begin{eqnarray} 
P(1,n)  & = {1\over q} \exp(\beta \mu) C^\gamma_n \exp(\beta \epsilon n) \exp(\beta \mu n) z^n & n \le \gamma_m  \\
& = {1 \over q}  C^\gamma_n  \exp(\beta \epsilon \gamma_m) \exp(\beta \mu n) z^n & n > \gamma_m 
\end{eqnarray} 

where $\beta \equiv 1/k_B T$, $q$ is the normalization, and $z$ is
introduced to account for the influence of the rest of the
lattice. The rest of the terms explicitly account for the interactions
and the chemical potential corresponding to the number of occupied
sites present.  Similarly, the probability that the center site is not
occupied, with $n$ neighbors occupied, is

\begin{equation} 
P(0,n) = {1\over q} C^\gamma_n \exp(\beta \mu n) z^n 
\end{equation} 

The normalization factor $q$ is given by the condition

\begin{equation} 
\sum_{n = 0}^{\gamma} \left[ P(1,n) + P(0,n)\right]  = 1 
\end{equation} 

The average occupancy  number $<n_0>$ for the 
the center site is the probability that the center is occupied
 regardless of the occupancy 
of the surrounding sites, that is: 

\begin{equation} 
<n_0> = \sum_{n = 0}^{\gamma} P(1,n).
\end{equation} 

The average occupancy  number $<n_e>$  for a given neighbor site is
 given by

\begin{equation} 
<n_e> = {1\over \gamma} \sum_{n = 0}^{\gamma} n \left[ P(1,n) + P(0,n)\right]. 
\end{equation} 

The unknown parameter $z$ is now determined by the condition that
these two occupancy  number, $<n_0>$ and $<n_e>$ are both equal
to $<n>$. Defining a function 

$$ g(z,\mu,T) = <n_0> - <n_e>, $$

the solution is found by the condition $g(z,\mu,T) = 0$. Determining
$z$ in this way, and using the expression for $<n_0>$ above, one obtains
the probability of occupation as a funciton of $T$ and $\mu$. In
addition, imposing the condition that the first and second derivatives 
of $g$ also vanish yeilds the condition for the critical point, since
this vanishing of the first and second derivatives marks the change
from the high temperature regime where only one solution exists, to
the presence of more than one solution at lower temperatures. In
addition, the condition that the first derivative vanishes is used to
locate the spinodal points.
Figure 3 shows the dependence of the critical density, temperature,
and the chemical potential on the parameter $\gamma_m$ which is varied
in integer steps for $\gamma = 12$. The case of $\gamma_m = \gamma$
reproduces the standard result for this approximation.  It is seen
that the critical temperature decreases very slowly at first as
$\gamma_m$ is reduced, but goes to zero for $\gamma_m =
1$. Interestingly, the case of $\gamma_m = 2$ displays a finite
critical temperature, which is somewhat surprising, since in this
case the system can form only two bonds at each site, and one expects
this to be equivalent to a one dimensional system with a vanishing
critical temperature.

\begin{figure}
\includegraphics[width=150mm]{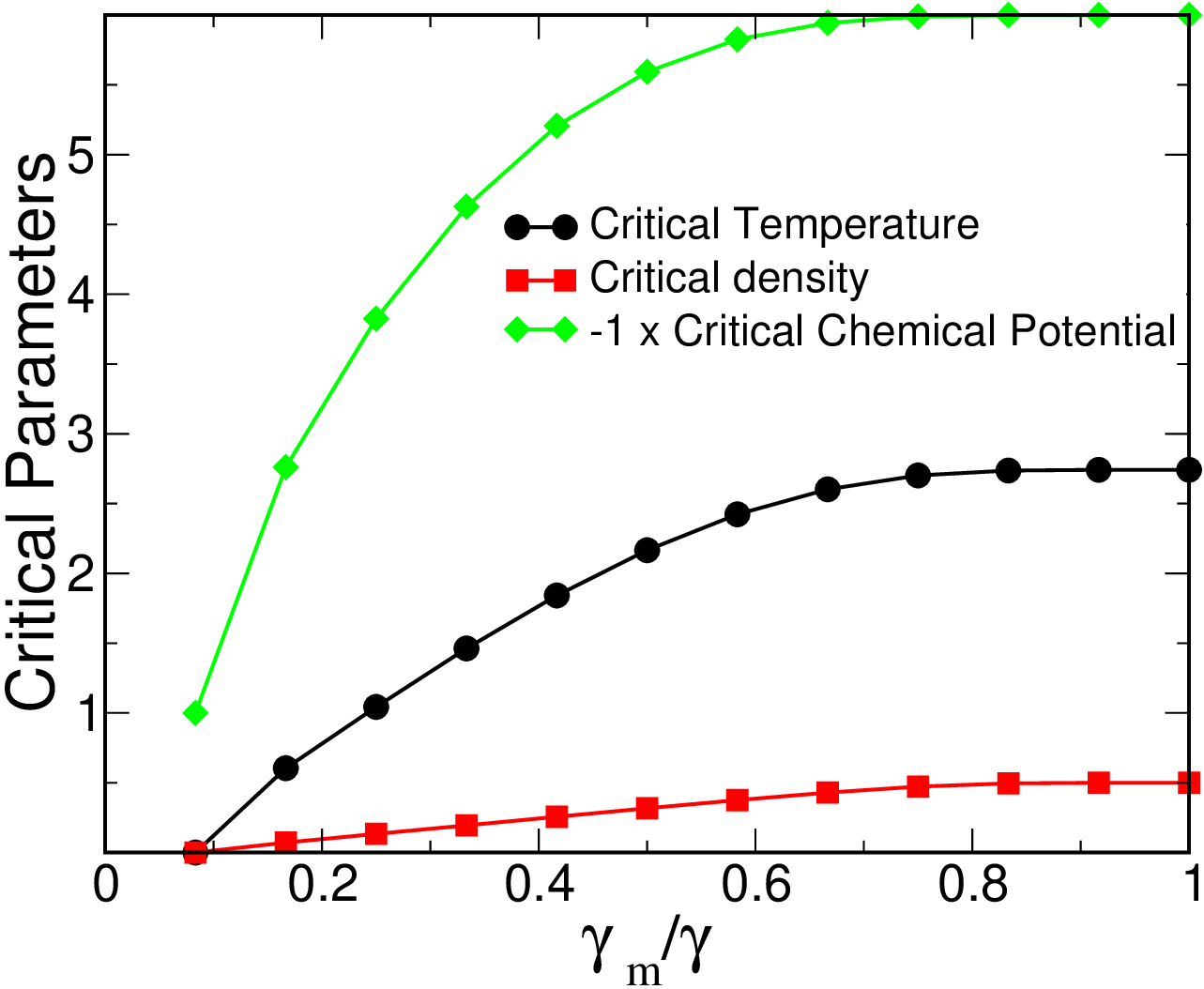}%
\caption{Critical density and critical temperature as a function of parameter 
$\gamma_m$ for the Bethe-Peierls approximation.}
\label{F_fig3}
\end{figure}

We next calculate the spinodals that demarcate the region of
instability in the $(<n>, T)$ phase diagram for $\gamma_m = 3$, which
is shown in Fig. 4. As in the mean field case the region of
instability also shrinks with the reduction in the valency.

\begin{figure}
\includegraphics[width=150mm]{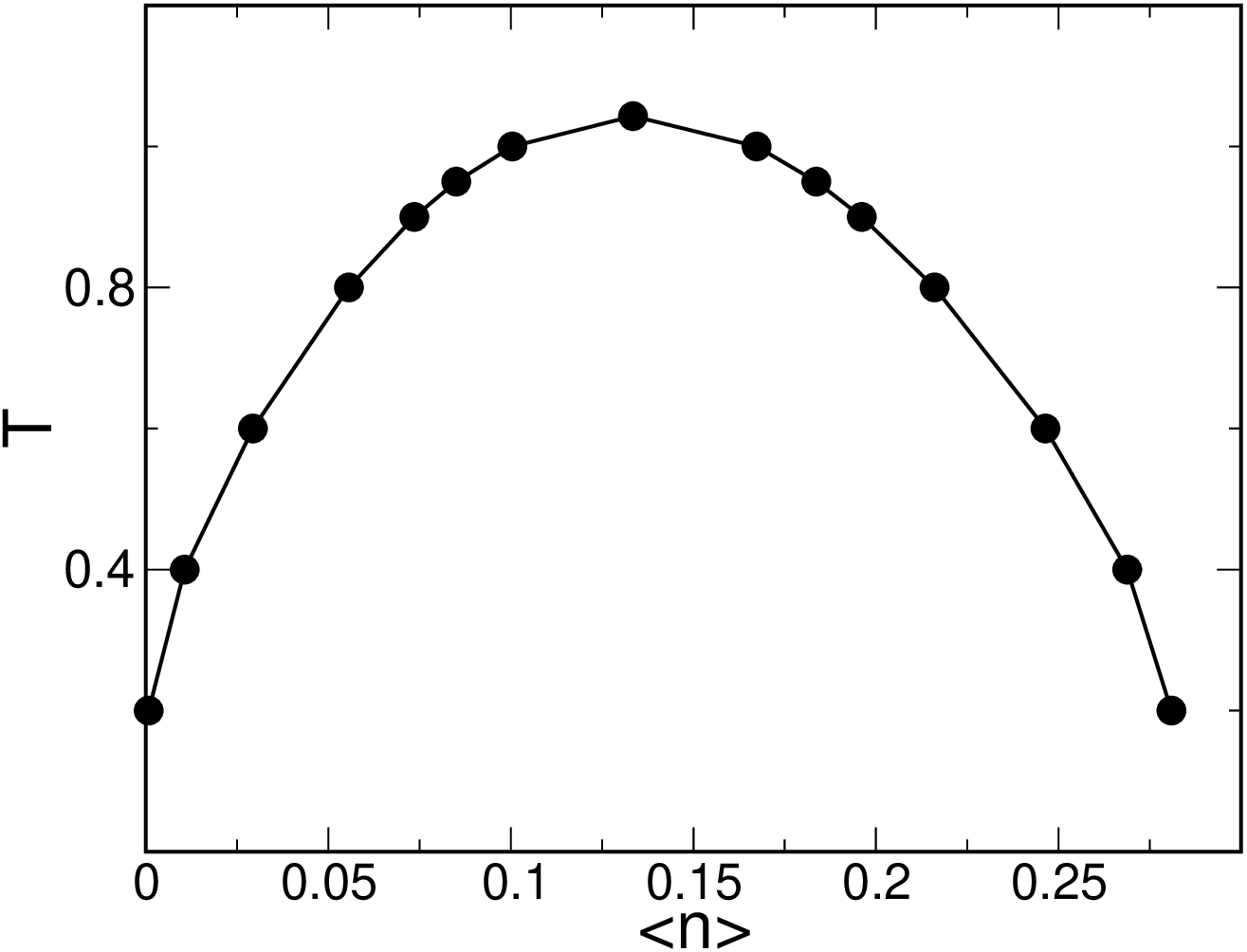}%
\caption{The  Bethe-Peierls spinodal line evaluated  for the specific case 
of $\gamma = 12$ and $\gamma_m = 3$ .}
\label{F_fig4}
\end{figure}

\section{Simulations} 

In order to estimate the accuracy of the approximations above, we
evaluate the critical parameters and the co-existence lines for
different values of $\gamma_m$ using computer simulations, emplying
the histogram reweighting technique
\cite{swen1,swen2,panagio1,panagio2}. In this method, histograms of
sampled energies and number densities from simulations at different
external parameters are used to estimate composite probability
distributions, which in turn may be used to obtain information on
phase equilibrium. To this end, we perform Monte Carlo simulations in
the constant temperature and constant chemical potential ensemble for
a lattice of $256$ sites arranged in an $FCC$ lattice. The energy of the 
system depends not only on which sites are occupied, but also which bonds 
exist in the system. Therefore, care must be taken that the formation and 
deletion of bonds obeys detailed balance. In the grand canonical simulations, 
it is sufficient, when a particle is inserted, to pick any of the possible bond configurations at random, with equal probability. At each
simulated temperature and chemical potential, runs of length $2.5$
million Monte Carlo cycles (MCS).  The energies and number of
particles every $100$ MCS, after an equilibration of $0.5$ million
MCS, are used to build histograms $f_i(N,E)$ (where $N$ is the number
of particles and $E$ is the energy), where $i = 1 \dots R$ indexes the
different runs at different $\mu$ and $T$ values. As described in
\cite{swen2,panagio2}, the composite probability distribtion
$\Gamma(N,E,\mu,\beta)$ is obtained with

\begin{equation}
\Gamma(N,E,\mu,\beta) = {\sum_{i=1}^R f_i (N,E) \exp(-\beta E + \beta \mu N) \over \sum_{i=1}^R K_i \exp(-\beta_i E + \beta_i \mu_i N -C_i) }
\end{equation} 
where $K_i$ is the total number of observations for each run, and the constants $C_i$ 
are obtained iteratively from the relationship

\begin{equation}
exp(C_i) = \sum_E \sum_N \Gamma(N,E,\mu_i,\beta_i).
\end{equation}

Once the above equations are run to convergence, the probability
distribution in particle number can be obtained by summing $\Gamma$ over
the energies. Phase co-existence at any given temperature is obtained
by requiring that the distribution with respect to density has two
distinct peaks, of equal height. Figure 5 shows the coexistence curves
obtained for the cases $\gamma_m = 12, 9, 6$. Studying phase
coexistence at lower values of $\gamma_m$ is severely hampered by the
extremely slow equilibration in the temperature range of interest and
has not been attempted. In each of the cases studied, evaluating the
coexisting phase densities is limited by the ability resolve the
coexisting density peaks, which is difficult as the critical point is
approached. Nevertheless, it is clear from the data shown that the
simulations confirm the trend seen in the calculations above, that the
critical tempeature, density and (minus) the chemical potential
decrease as $\gamma_m$ decreases. Figure 6 shows the critical
 parameters in comparison with the Bethe-Peierls calculations. It is
seen that the critical paramters drop more rapidly than suggested by
the Bethe-Peierls calculations. Unlike the $\gamma_m = 12$ case, 
where the coexistence chemical potential and the mean of the coexisting 
densities are constant below the critical temperature, both these quantities
increase as one moves to lower temperatures below the critical temperature.

\begin{figure}
\includegraphics[width=150mm]{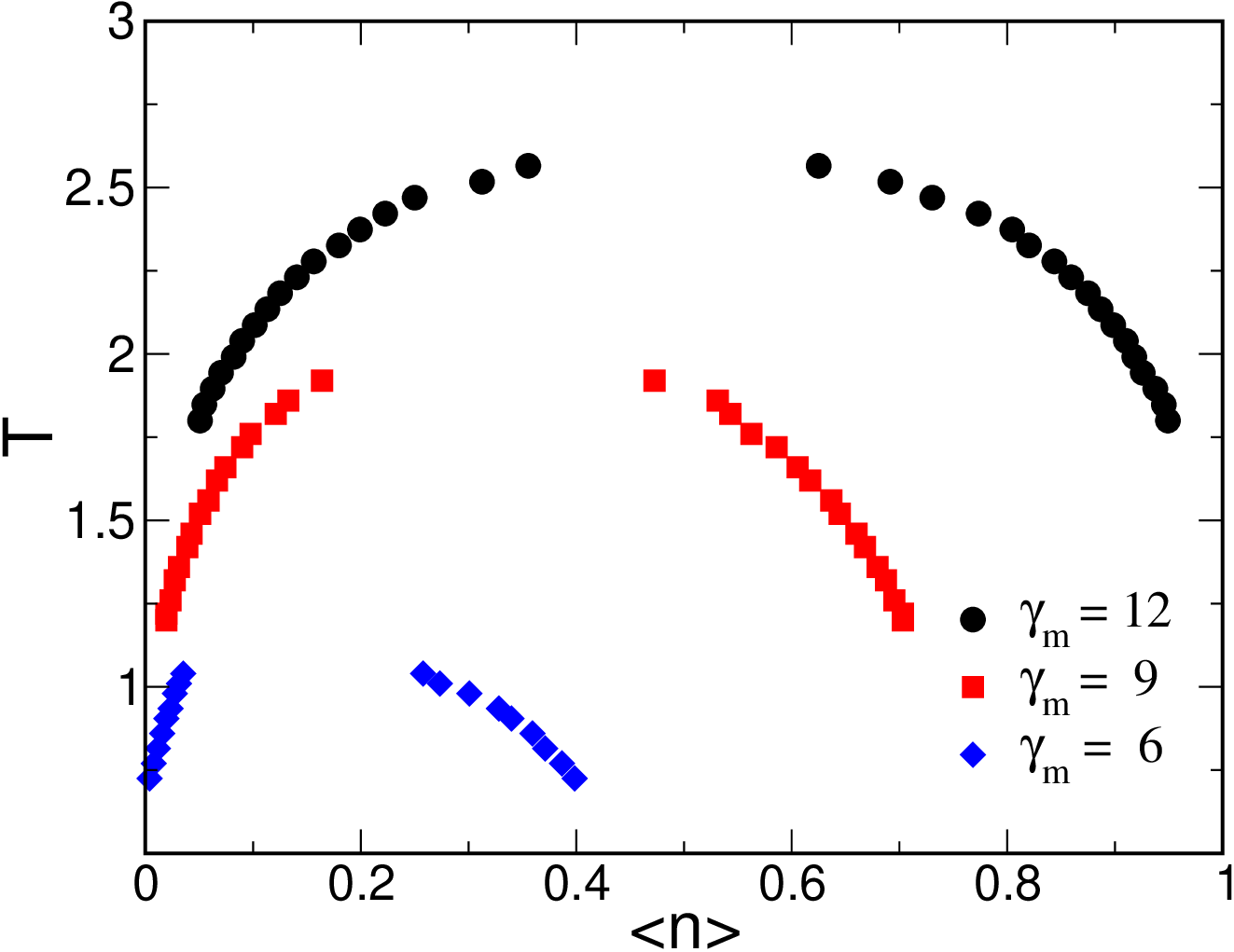}%
\caption{The coexistence curves for $\gamma_m = 12, 9, 6$ from simulations, 
showing that the coexistence region shrinks as $\gamma_m$ decreases.}
\label{F_fig5}
\end{figure}

\begin{figure}
\includegraphics[width=150mm]{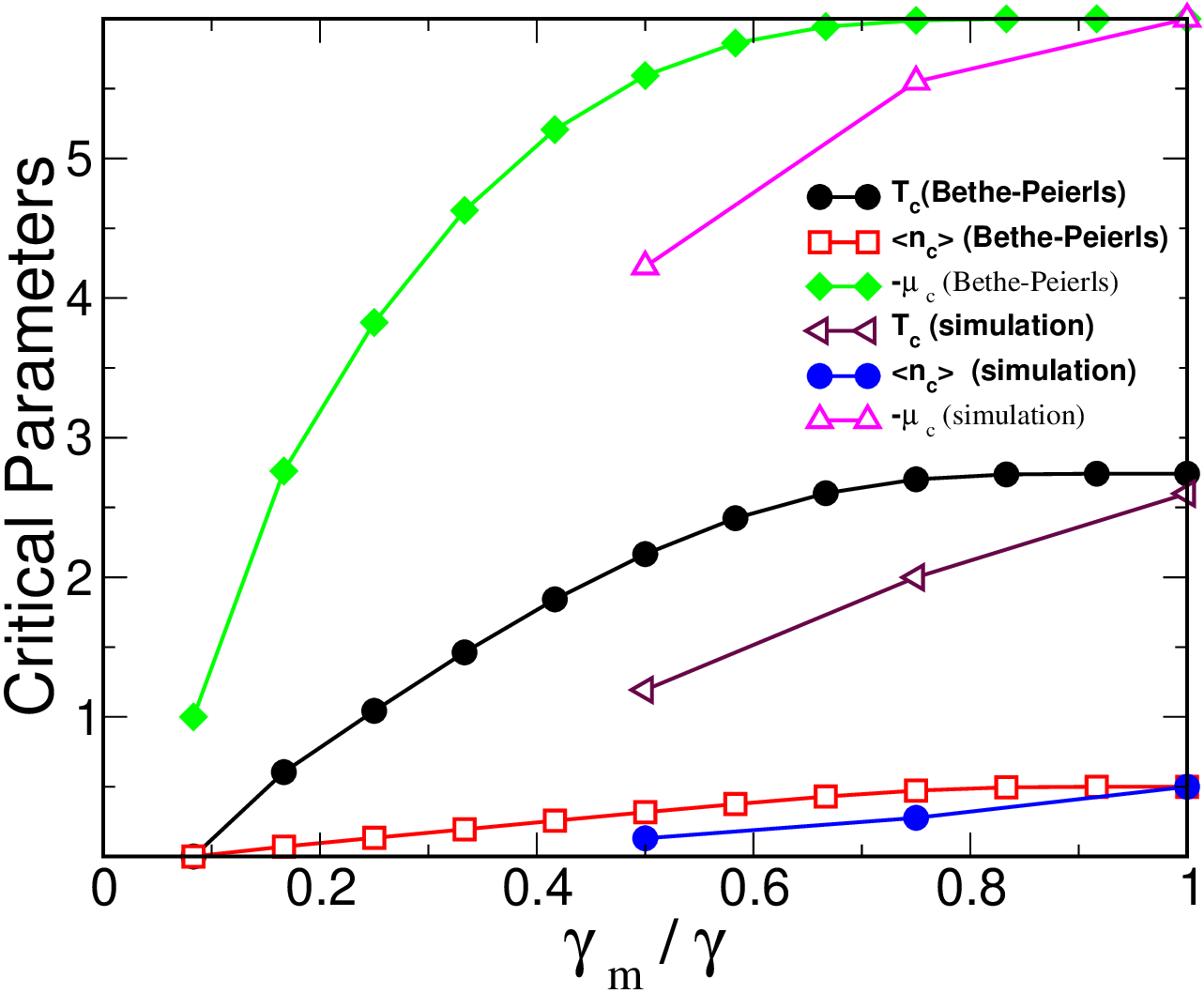}%
\caption{Comparison of the critical parameters from simulation with those of the Bethe-Peierls approximation, 
indicating that the critical parameters decrease more rapidly than suggested by the Bethe-Peierls approximation.}
\label{F_fig6}
\end{figure}
 
\section{Conclusions} 

We have presented results for lattice gas models with the imposition
of a constraint on the maximum number of bonds that particles can
participate in. Mean field (Bragg-Williams and Bethe-Peierls)
calculations and computer simulations using the histogram reweighting
technique, show that the critical temperature and density decrease as
the maximum number of bonds allowed for a given particle is
reduced. Also, we find that the density range of the coexistence
region, where the liquid and gas phases coexist, shrinks as the
maximum valency of the particles is reduced.  These results are
consistent with the results that have been obtained earlier for
similarly defined continuum models \cite{Zac05b,bianchi2006}. It is
thus confirmed that valency reduction is effective in opening a large
region of densities in which low temperature disordered states can be
accessed in equilibrium.  Studies of models with controlled valency,
both on the lattice and off-lattice, may make it possible to
disentangle gelation from phase separation.

 \end{document}